\documentclass[aps,showkeys,graphicx,twocolumn]{revtex4}
\usepackage{amssymb}
\usepackage{amsmath}
\usepackage{graphicx}
\usepackage{array}
\usepackage{color}
\usepackage{epstopdf}

\begin{document}

\title{Quantum entanglement generation on magnons assisted with microwave cavities coupled to a superconducting qubit}

\author{Jiu-Ming Li}
\author{Shao-Ming Fei}\email{feishm@cnu.edu.cn}
\affiliation{School of Mathematical Sciences, Capital Normal University, Beijing 100048, China}

\begin{abstract}
We present protocols to generate quantum entanglement on nonlocal magnons in hybrid systems composed of yttrium iron garnet (YIG) spheres, microwave cavities and a superconducting (SC) qubit. In the schemes, the YIGs are coupled to respective microwave cavities in resonant way, and the SC qubit is placed at the center of the cavities, which interacts with the cavities simultaneously. By exchanging the virtual photon, the cavities can indirectly interact in the far-detuning regime. Detailed protocols are presented to establish entanglement for two, three and arbitrary $N$ magnons with reasonable fidelities.
\end{abstract}



\maketitle

\section{INTRODUCTION}

Quantum entanglement is one of the most important features in quantum mechanics. The quantum entangled states \cite{EPR,GHZ1,W1,QE} are significant ingredients in quantum information processing. Over past decades, various theoretical and experimental proposals have been presented for processing quantum information by using various systems such as atoms \cite{AP,A1,A2,A3,A4,A5,A6,A7,A8,A9}, spins \cite{S1,S2,S3,S4,S5,S6,S7}, ions \cite{I1,I2,I3,I4,I5,I6,I7,I8}, photons \cite{AP,PT1,PT2,PT3,PT4,PT5,PT6,PT7,PT8,PT9,PT10}, phonons \cite{PN1,PN2,PN3}, and so on. With the development of technologies, the quantum entanglement has been established not only in microscopic systems, but also in the macroscopic systems such as superconducting circuits \cite{SC1,SC2,SC3,SC4,SC5,SC6} and magnons system \cite{LJ1,LJ2,Kong,M1,M2,M3}.

Hybrid systems exploit the advantages of different quantum systems in achieving certain quantum tasks, such as creating quantum entanglement and carrying out quantum logic gates. Many works have been presented so far for quantum information processing in the hybrid systems \cite{HS1,HS2,HS3,HS4}. For instance, as an important quantum technology \cite{HSC}, the hybrid quantum circuits combine superconducting systems with other physical systems which can be fabricated on a chip. The superconducting (SC) qubit circuits \cite{SCQ1,SCQ2}, based on the Josephson junctions, can exhibit quantum behaviors even at macroscopic scale. Generally, the interaction between the SC qubits and the environment, e.g., systems in strong or even ultrastrong coupling regime via quantized electromagnetic fields, would result in short coherence time. Thus many researches on circuit quantum electrodynamics (QED) \cite{CQED} have been presented with respect to the SC qubits, superconducting coplanar waveguide resonators, \emph{LC} resonators and so on. This circuit QED focuses on studies of the light-matter interaction by using the microwave photons, and has become a relative independent research field originated from cavity QED.

The hybrid systems composed of collective spins (magnons) in ferrimagnetic systems and other systems are able to constitute the magnon-photon \cite{MT1,MT2}, magnon-phonon \cite{MN3,MN1,MN2}, magnon-photon-phonon \cite{LJ1,LJ2,MTN} systems and so on, giving rise to new interesting applications. Ferrimagnetic systems such as yttrium iron garnet (YIG) sphere have attracted considerable attention in recent years, which provide new platforms for investigating the macroscopic quantum phenomena particularly. Such systems are able to achieve strong and even ultrastrong couplings \cite{SU} between the magnons and the microwave photons, as a result of the high density of the collective spins in YIG and the lower dissipation. The YIG has the unique dielectric microwave properties with very lower microwave magnetic loss parameter. Meanwhile, some important works have been presented on magnon Kerr effect \cite{Kerr1,Kerr2}, quantum transduction \cite{MQT}, magnon squeezing \cite{MS1,MS2}, magnon Fock state \cite{MFS} and entanglement of magnons. For example, In 2018 Li \emph{et al}. \cite{LJ1} proposed a system consisted of magnons, microwave photons and phonons for establishing tripartite entangled states based on the magnetostrictive interaction and that the entangled state in magnon-photon-phonon system is robust. In 2019 Li \emph{et al.} \cite{LJ2} constructed the entangled state of two magnon modes in a cavity magnomechanical system by applying a strong red-detuned microwave field on a magnon mode to activate the nonlinear magnetostrictive interaction. In 2021 Kong \emph{et al.} \cite{Kong} used the indirect coherent interaction for accomplishing two magnons entanglement and squeezing via virtual photons in the ferromagnetic-superconducting system.

In this work, we first present a hybrid system composed of two YIG spheres, two identical microwave cavities and a SC qubit to establish quantum entanglement on two nonlocal magnons. In this system, two YIGs are coupled to respective microwave cavities that cross each other. And a SC qubit is placed at the center of the crossing of two identical cavities, namely, the SC qubit interacts with the two cavities simultaneously. The magnons in YIGs can be coupled to the microwave cavities in the resonant way, owing to that the frequencies of two magnons can be tuned by biased magnetic fields, respectively. Compared with other works, the SC qubit is coupled to the two microwave cavities in the far-detuning regime, meaning that the two identical cavities indirectly interact with each other by exchanging virtual photons. Then, we give the effective Hamiltonian of the subsystem composed of the SC qubit and two cavities, and present the protocol of entanglement establishment. In Sec. \ref{3M}, we consider the case of three magnons. In the hybrid system shown in Fig.\ref{fig3}, the three identical microwave cavities could indirectly interact via the virtual photons, and each magnon is resonant with the respective cavity by tuning the frequency of the magnon. At last, we get the isoprobability entanglement on three nonlocal magnons. Moreover, the hybrid system composed of \emph{N} magnons, \emph{N} identical microwave cavities and a SC qubit is derived in Sec. \ref{4M}. We summarize in Sec. \ref{5M}.

\section{QUANTUM ENTANGLEMENT ON TWO NONLOCAL MAGNONS}\label{2M}
\subsection{Hamiltonian of the hybrid system}

\begin{figure}
\centering
\includegraphics[width=8.5cm,angle=0]{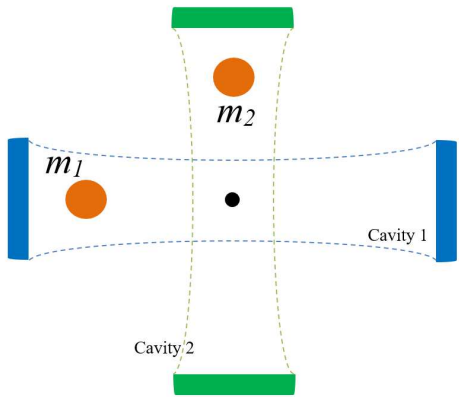}
\caption{(Color online) Schematic of the hybrid system composed of two yttrium iron garnet spheres coupled to respective microwave cavities. Two cavities cross each other, and a superconducting qubit (black spot) is placed at the center of the crossing. }\label{fig1}
\end{figure}

We consider a hybrid system, see Fig.\ref{fig1}, in which two microwave cavities cross each other, two yttrium iron garnet (YIG) spheres are coupled to the microwave cavities, respectively. A superconducting (SC) qubit, represented by black spot in the Fig.\ref{fig1}, is placed at the center of the crossing in order to interact with the two microwave cavities simultaneously. The YIG spheres are placed at the antinode of two microwave magnetic fields, respectively, and a static magnetic field is locally biased in each YIG sphere. In our model, the SC qubit is a two-level system with ground state $|g\rangle\!_q$ and excited state $|e\rangle\!_q$.

The magnetostatic modes in YIG can be excited when the magnetic component of the microwave cavity field is perpendicular to the biased magnetic field. We only consider the Kittel mode \cite{Kittel} in the hybrid system, namely, the magnon modes can be excited in YIG. The frequency of the magnon is in the gigahertz range. Thus the magnon generally interacts with the microwave photon via the magnetic dipole interaction. The frequency of the magnon is given by $\omega_m=\gamma H$, where $H$ is the biased magnetic field and $\gamma/2\pi=28$ GHz/T is the gyromagnetic ratio.

In recent years, some experiments have already realized the strong and ultrastrong magnon-magnon coupling \cite{mm1,mm2,mm3} as well as the magnon-qubit interaction \cite{MSC,PMS}, which means that in the hybrid system shown in Fig.\ref{fig1} the magnon is both coupled to the SC qubit and another magnon. However, we mainly consider that the magnons which frequencies are tuned by the locally biased static magnetic fields can be resonant with the cavities. In the meantime, the two cavities modes interact indirectly in the far-detuning regime for exchanging photons. The entanglement of two nonlocal magnons can be constructed by using two cavities and the SC qubit. Given that there are magnon-magnon and magnon-qubit interactions, the magnon can be detuned with the qubit and another magnon in order to neglect their interactions. In the rotating wave approximation the Hamiltonian of the hybrid system is ($\hbar=1$ hereafter) \cite{Hm}
\begin{eqnarray}     
H^{(\mathrm{S})}\!=H_0+H_{\mathrm{int}}  \nonumber
\end{eqnarray}
\begin{eqnarray}
H_0&\!\!\!=\!\!\!&\omega_{m_1}m_1^\dag m_1+\omega_{m_2}m_2^\dag m_2+\frac{1}{2}\omega_q\sigma_z    \nonumber    \\
&&+\omega_{a_1}a_1^\dag a_1+\omega_{a_2}a_2^\dag a_2   \nonumber
\end{eqnarray}
\begin{eqnarray}
H_{\mathrm{int}}&\!\!\!=\!\!\!&\lambda_{m_1}(a_1m^\dag_1+a_1^\dag m_1)+\lambda_{m_2}(a_2m^\dag_2+a_2^\dag m_2)     \nonumber     \\
&&+\lambda_{q_1}(a_1\sigma^++a_1^\dag\sigma)+\lambda_{q_2}(a_2\sigma^++a_2^\dag\sigma).
\end{eqnarray}
Here, $H_0$ is the free Hamiltonian of the two cavities, two magnons and the SC qubit. $H_{\mathrm{int}}$ is the interaction Hamiltonian among the cavities, magnons and SC qubit. $\omega_{m_1}$ and $\omega_{m_2}$ are the frequencies of the two magnons, which are tunable under biased magnetic fields, respectively. $\omega_{a_1}$ and $\omega_{a_2}$ are the frequencies of two cavities, and $\omega_q$ is the state transition frequency between $|g\rangle\!_q\leftrightarrow|e\rangle\!_q$ of the SC qubit. In the Kittel mode, the collective spins in YIGs can be expressed by the boson operators. $m_1$ ($m_2$) and $m_1^\dag$ ($m_2^\dag$) are the annihilation and creation operators of magnon mode 1 (2). $a_1$ ($a_2$) and $a_1^\dag$ ($a_2^\dag$) denote the annihilation and creation operators of cavity mode 1 (2), respectively. They satisfy commutation relations $[O,O^\dag]=1$ for $O=a_1, a_2, m_1, m_2$. $\sigma_z=|e\rangle_q\langle e|-|g\rangle_q\langle g|$. $\sigma=|g\rangle_q\langle e|$ and $\sigma^+=|e\rangle_q\langle g|$ are the lowing and raising operators of the SC qubit. $\lambda_{q_1}$ ($\lambda_{q_2}$) is the coupling strength between the SC qubit and the cavity mode 1 (2). $\lambda_{m_1}$ ($\lambda_{m_2}$) is the coupling between the magnon mode 1 (2) and the cavity mode 1 (2).

As mentioned above, the two microwave cavities are identical ones with the same frequency $\omega_{a_1}=\omega_{a_2}=\omega_a$. Meanwhile, one can assume that $\lambda_{q_1}=\lambda_{q_2}=\lambda_q$. In the interaction picture with respect to $e^{-\mathrm{i}H_0 t}$, the Hamiltonian is expressed as
\begin{eqnarray}    
H^{(\mathrm{I})}&\!\!=\!\!&\lambda_{m_1}a_1m^\dag_1e^{\mathrm{i}\delta_1 t}+\lambda_{m_2}a_2m^\dag_2e^{\mathrm{i}\delta_2 t}+\lambda_qa_1\sigma^+e^{\mathrm{i}\Delta_1 t} \nonumber    \\
&&+\lambda_qa_2\sigma^+e^{\mathrm{i}\Delta_2 t}+H.c.,
\end{eqnarray}
where $\delta_1=\omega_{m_1}-\omega_a$, $\delta_2=\omega_{m_2}-\omega_a$, $\Delta_1=\omega_q-\omega_a$ and $\Delta_2=\omega_q-\omega_a$. The SC qubit is coupled to the two cavities simultaneously. Owing to $\Delta_1=\Delta_2=\Delta_0\neq0$ and $\Delta_0\gg\lambda_q$, the two identical microwave cavities indirectly interact with each other in the far-detuning regime. Therefore, the effective Hamiltonian of the subsystem composed of the two microwave cavities and the SC qubit in the far-detuning regime is given by \cite{EH}
\begin{eqnarray}    
H_{\mathrm{eff}}\!=\!\widetilde{\lambda}_q\!\left[\sigma_z(a_1^{\dag}a_1+a_2^{\dag}a_2+a_1^{\dag}a_2+a_1a_2^{\dag})+2|e\rangle_q\langle e|\right], \label{eff1}
\end{eqnarray}
where $\widetilde{\lambda}_q=\lambda_q^2/\Delta_0$.

\subsection{Entangled state generation on two nonlocal magnons}

We now give the protocol of quantum entanglement generation on two nonlocal magnons. Generally, the magnon can be excited by a drive magnetic field. For convenience the state of magnon 1 is prepared as $|1\rangle\!_{m\!_1}$ via the magnetic field. The initial state of the hybrid system is $|\varphi\rangle_0=|1\rangle\!_{m\!_1}|0\rangle\!_{m\!_2}|0
\rangle\!_{a\!_1}|0\rangle\!_{a\!_2}|g\rangle\!_q$, in which the two cavities are all in the vacuum state, magnon 2 is in the state $|0\rangle\!_{m\!_2}$, and the SC qubit is in state $|g\rangle\!_q$ which is unaltered all the time due to the indirect interaction between the two cavities.

\emph{step 1}: The frequency of magnon 1 is tuned to be $\omega_{m_1}\!=\!\omega_{a_1}$ so that the cavity 1 could be resonated with it. Therefore, the magnon 1 and cavity 1 are in a superposed state after time $T_1=\pi/4\lambda_{m_1}$. The local evolution is $|1\rangle\!_{m\!_1}|0\rangle\!_{a\!_1}\rightarrow\frac{1}{\sqrt{2}}(|1\rangle\!_{m\!_1}|0\rangle\!_{a\!_1}-\mathrm{i}|0\rangle\!_{m\!_1}|1\rangle\!_{a\!_1})$, which means that the states of SC qubit, magnon 2 and cavity 2 are unchanged due to decoupling between the SC and two cavities, and the magnon 2 is far-detuned with cavity 2. The state evolves to
\begin{eqnarray}     
|\varphi\rangle_1&\!\!\!=\!\!\!&\frac{1}{\sqrt{2}}(|1\rangle\!_{m\!_1}|0\rangle\!_{a\!_1}-\mathrm{i}|0\rangle\!_{m\!_1}|1\rangle\!_{a\!_1})   \nonumber   \\
&&\otimes|0\rangle\!_{m\!_2}\otimes|0\rangle\!_{a\!_2}\otimes|g\rangle\!_q.
\end{eqnarray}

\emph{step 2}: The magnons are tuned to far detune with respective cavities. From Eq. (\ref{eff1}), the evolution of subsystem composed of two microwave cavities and SC qubit is given by
\begin{eqnarray}     
|\chi(t)\rangle\!_{\mathrm{sub}}&\!\!=\!\!&e^{\mathrm{i}\widetilde{\lambda}_qt}\big[\cos(\widetilde{\lambda}_qt)|1\rangle\!_{a\!_1}|0\rangle\!_{a\!_2}
+\mathrm{i}\sin(\widetilde{\lambda}_qt)|0\rangle\!_{a\!_1}|1\rangle\!_{a\!_2}\big]     \nonumber   \\
&&\otimes|g\rangle\!_q
\end{eqnarray}
under the condition $\Delta_0\gg\lambda_q$.

After time $T_2=\pi/2\widetilde{\lambda}_q$, the evolution between two cavities is $|1\rangle\!_{a\!_1}|0\rangle\!_{a\!_2}\rightarrow-|0\rangle\!_{a\!_1}|1\rangle\!_{a\!_2}$, which indicates that the photon can be indirectly transmitted between the two cavities, with the state of SC qubit unchanged. Therefore, the state after this step changes to
\begin{eqnarray}      
|\varphi\rangle_2&\!\!\!=\!\!\!&\frac{1}{\sqrt{2}}(|1\rangle\!_{m\!_1}|0\rangle\!_{a\!_1}|0\rangle\!_{a\!_2}
+\mathrm{i}|0\rangle\!_{m\!_1}|0\rangle\!_{a\!_1}|1\rangle\!_{a\!_2})  \nonumber  \\
&&\otimes|0\rangle\!_{m\!_2}\otimes|g\rangle\!_q.
\end{eqnarray}

\emph{step 3}: The frequency of magnon 2 is tuned with $\omega_{m_2}=\omega_{a_2}$ to resonate with the cavity 2. In the meantime the cavities are decoupled to the SC qubit and the magnon 1 is far detuned with the cavity 1. After time $T_3=\pi/2\lambda_{m_2}$, the local evolution $|0\rangle\!_{m\!_2}|1\rangle\!_{a\!_2}\rightarrow-\mathrm{i}|1\rangle\!_{m\!_2}|0\rangle\!_{a\!_2}$ is attained. The final state is
\begin{eqnarray}      
|\varphi\rangle_3&\!\!\!=\!\!\!&\frac{1}{\sqrt{2}}(|1\rangle\!_{m\!_1}|0\rangle\!_{m\!_2}
+|0\rangle\!_{m\!_1}|1\rangle\!_{m\!_2})    \nonumber   \\
&&\otimes|0\rangle\!_{a\!_1}\otimes|0\rangle\!_{a\!_2}\otimes|g\rangle\!_q,
\end{eqnarray}
which is just the single-excitation Bell state on two nonlocal magnons.

In the whole process, we mainly consider the interactions between the magnons and the cavities, and between the cavities and the SC qubit. However, the SC qubit can be coupled to the magnons. In terms of Ref.\cite{MSC}, the interactions between the magnons and the SC qubit are described as $H_{qm,1}=\lambda_{qm,1}(\sigma^+m_1+H.c.)$ and $H_{qm,2}=\lambda_{qm,2}(\sigma^+m_2+H.c.)$ where $\lambda_{qm,1}=\lambda_q\lambda_{m_1}/\Delta_0$ and $\lambda_{qm,2}=\lambda_q\lambda_{m_2}/\Delta_0$, while the conditions $\omega_q=\omega_{m_1}$ and $\omega_q=\omega_{m_2}$ are attained. In the meantime, the two magnons are interacts each other by using the SC qubit. Generally, the frequencies of two magnon modes are tuned by the locally biased magnetic fields. Therefore, the magnon can be detuned with the SC qubit and another magnon in order to neglect the interactions between the magnons and the SC qubit.

\subsection{Numerical result}

\begin{figure*}
\centering
\begin{tabular}{cc}
\includegraphics[width=9cm,angle=0]{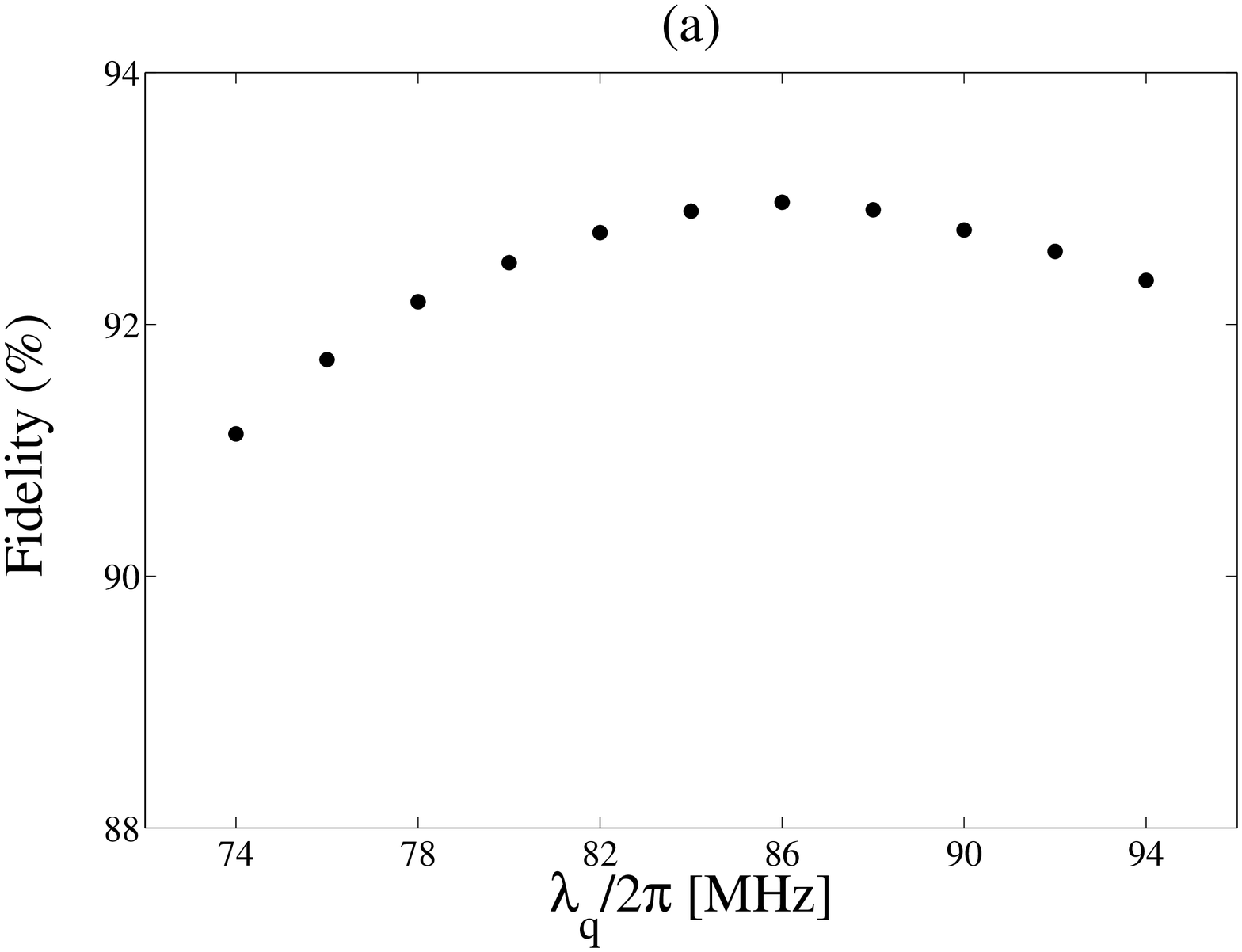} & \includegraphics[width=9cm,angle=0]{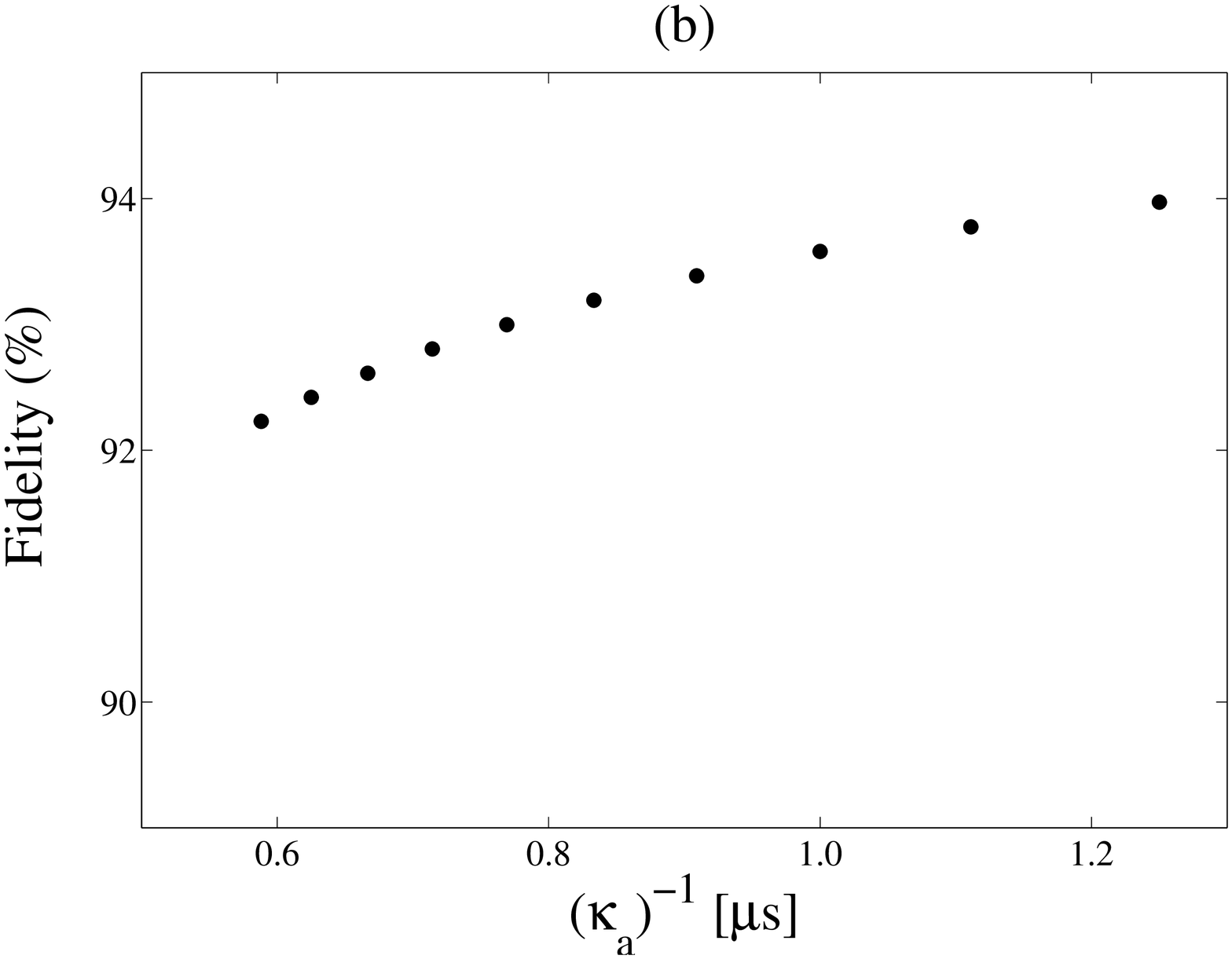}\\
\includegraphics[width=9cm,angle=0]{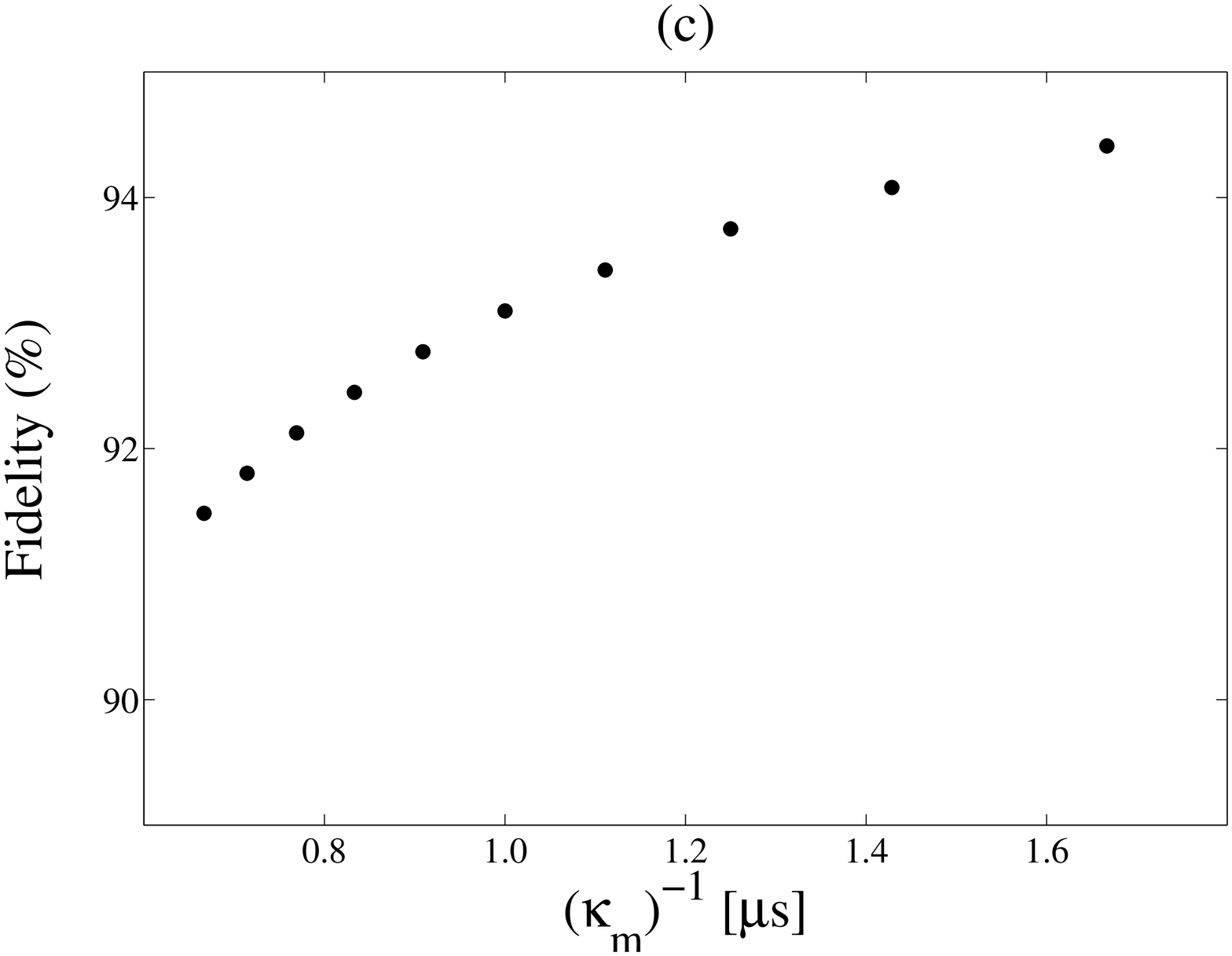} & \includegraphics[width=9cm,angle=0]{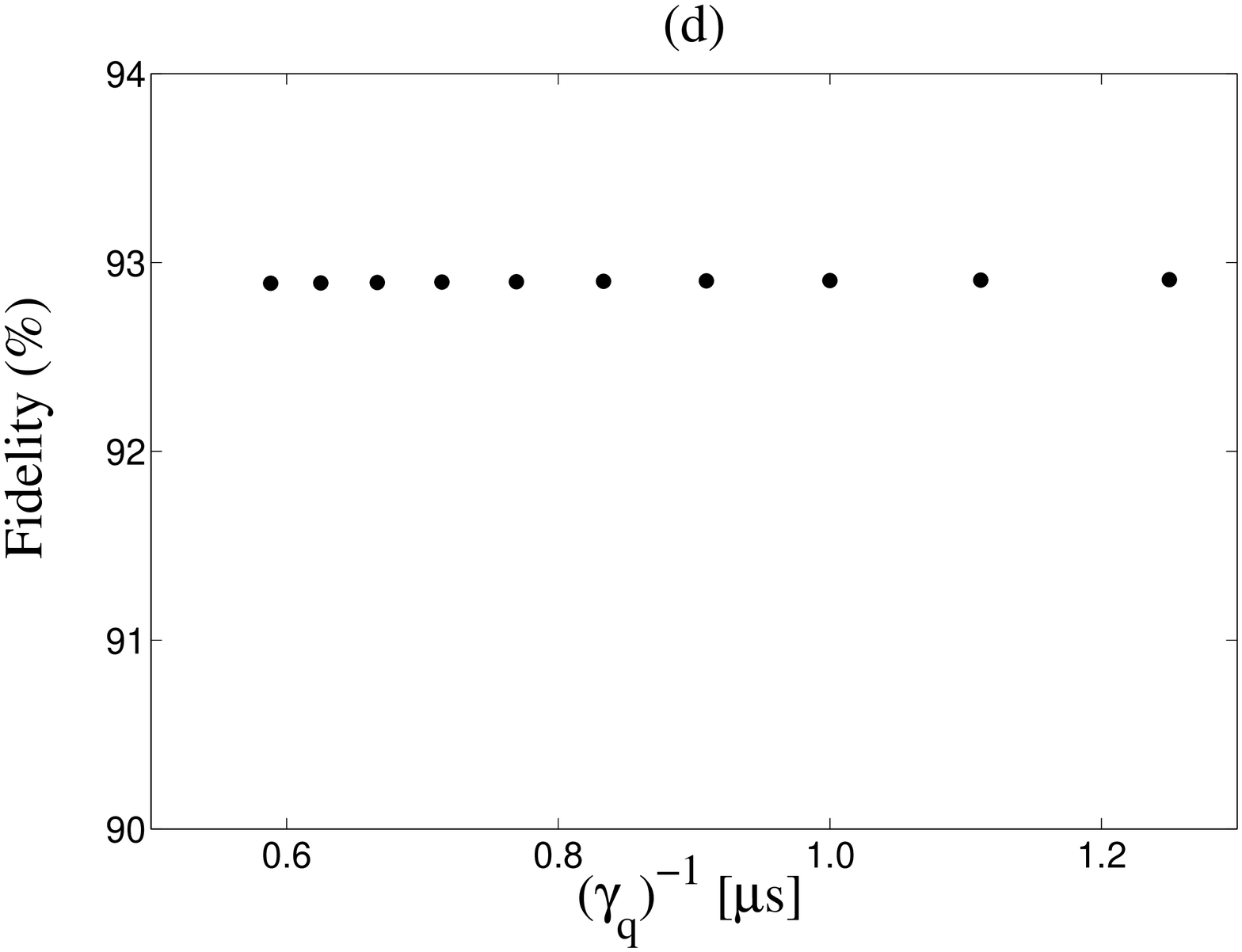}
\end{tabular}
\caption{(a) The fidelity of the Bell state of two nonlocal magnons with respect to the coupling strength $\lambda_q$. Since $\widetilde{\lambda}_q=\lambda_q^2/\Delta_0$ in Eq.(\ref{eff1}), the fidelity is similar to parabola. (b)-(d) The fidelity of the Bell state versus the dissipations of cavities, magnons, and SC qubit, respectively. }\label{fig2}
\end{figure*}

We here simulate \cite{SIM} the fidelity of the Bell state on two nonlocal magnons by considering the dissipations of all constituents of the hybrid system. The realistic evolution of the hybrid system composed of magnons, microwave cavities and SC qubit is governed by the master equation
\begin{eqnarray}    
\dot{\rho}&\!\!=\!\!&-\mathrm{i}[H^{(I)},\rho]+\kappa_{m_1}D[m_1]\rho+\kappa_{m_2}D[m_2]\rho\nonumber   \\
&&+\kappa_{a_1}D[a_1]\rho+\kappa_{a_2}D[a_2]\rho+\gamma_qD[\sigma]\rho.
\end{eqnarray}
Here, $\rho$ is the density operator of the hybrid system, $\kappa_{m_1}$ and $\kappa_{m_2}$ are the dissipation rates of magnon 1 and 2, $\kappa_{a_1}$ and $\kappa_{a_2}$ denote the dissipation rates for the two microwave cavities 1 and 2, $\gamma_q$ is the dissipation rate of the SC qubit, $D[X]\rho\!\!=\!\!(2X\rho X^\dag-X^\dag X\rho-\rho X^\dag X)/2$ for $X=m_1, m_2, a_1, a_2, \sigma$. The fidelity of the entangled state of two nonlocal magnons is defined by $F=_3\!\!\langle\varphi|\rho|\varphi\rangle_3$.

The related parameters are chosen as $\omega_q/2\pi=7.92$ GHz, $\omega_a/2\pi=6.98$ GHz, $\lambda_q/2\pi=83.2$  MHz, $\lambda_{m_1}/2\pi=15.3$ MHz, $\lambda_{m_2}/2\pi=15.3$ MHz \cite{PMS}, $\kappa_{m_1}/2\pi$=$\kappa_{m_2}/2\pi=\kappa_m/2\pi=1.06$ MHz, $\kappa_{a_1}/2\pi=\kappa_{a_2}/2\pi=\kappa_a/2\pi=1.35$ MHz \cite{SU}, $\gamma_q/2\pi=1.2$ MHz \cite{MSC}. The fidelity of the entanglement between two nonlocal magnons can reach 92.9$\%$.

The influences of the imperfect relationship among parameters is discussed next. The Fig.\ref{fig2}(a) shows the fidelity influenced by the coupling strength between the microwave cavities and the SC qubit. Since $\widetilde{\lambda}_q=\lambda_q^2/\Delta_0$ in Eq.(\ref{eff1}), the fidelity is similar to parabola. In Fig.\ref{fig2}(b)-(d), we give the fidelity varied by the dissipations of cavities, magnons, and SC qubit. As a result of the virtual photon, the fidelity is almost unaffected by the SC qubit, shown in Fig.\ref{fig2}(d).

\section{ENTANGLEMENT GENERATION FOR THREE NONLOCAL MAGNONS}\label{3M}
\subsection{Entangled state of three nonlocal magnons}

\begin{figure}
\centering
\includegraphics[width=8.5cm,angle=0]{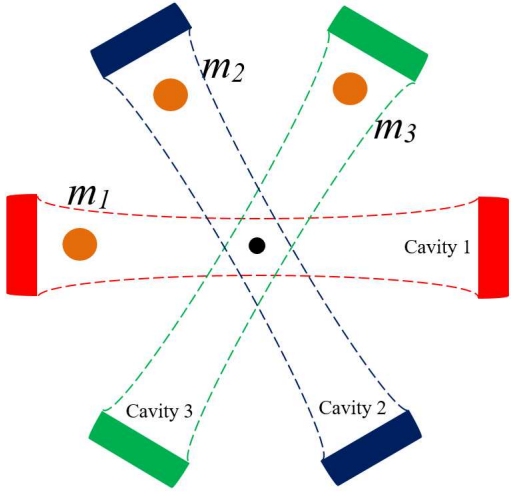}
\caption{(Color online) Schematic of the hybrid system composed of three yttrium iron garnet spheres coupled to respective microwave cavities. A superconducting qubit (black spot) is placed at the center of the three cavities. }\label{fig3}
\end{figure}

Similar to the protocol of entangled state generation for two nonlocal magnons in two microwave cavities, we consider the protocol for entanglement of three nonlocal magnons. As shown in Fig.\ref{fig3}, similar to the hybrid system composed of two magnons coupled to the respective microwave cavities and a SC qubit in Fig.\ref{fig1}, there are three magnons in three YIGs coupled to respective microwave cavities and a SC qubit placed at the center of the three identical cavities ($\omega_{a_1}\!=\!\omega_{a_2}\!=\!\omega_{a_3}\!=\!\omega_a$). Each magnon is in biased static magnetic field and is located at the antinode of the microwave magnetic field.

In the interaction picture, the Hamiltonian of the hybrid system depicted in Fig.\ref{fig3} is
\begin{eqnarray}     
H^{(\mathrm{I})}_3&\!\!=\!\!&\lambda_{m_1}a_1m^\dag_1e^{\mathrm{i}\delta_1 t}+\lambda_{m_2}a_2m^\dag_2e^{\mathrm{i}\delta_2 t}     \nonumber     \\
&&+\lambda_{m_3}a_3m^\dag_3e^{\mathrm{i}\delta_3 t}+\lambda_qa_1\sigma^+e^{\mathrm{i}\Delta_1 t}        \nonumber    \\
&&+\lambda_qa_2\sigma^+e^{\mathrm{i}\Delta_2 t}+\lambda_qa_3\sigma^+e^{\mathrm{i}\Delta_3 t}+H.c.,
\end{eqnarray}
where $\lambda_{m_3}$ is the coupling strength between magnon 3 and microwave cavity 3, $a_3$ and $m^\dag_3$ are annihilation operator of the cavity 3 and creation operator of the magnon 3, respectively. $\lambda_q$ is the coupling between the SC qubit and three cavities, $\delta_3=\omega_{m_3}-\omega_a$. The frequency $\omega_{m_3}$ can be tuned by the biased magnetic field in microwave cavity 3. $\Delta_3=\omega_q-\omega_a=\Delta_0$.

At the beginning we have the initial state $|\psi\rangle_0^{(3)}\!\!\!=\!\!|\psi\rangle\!_m^{(3)}\otimes|\psi\rangle\!_a^{(3)}\otimes|g\rangle\!_q$ with
$|\psi\rangle\!_m^{(3)}\!\!\!=\!\!|1\rangle\!_{m\!_1}|0\rangle\!_{m\!_2}|0\rangle\!_{m\!_3}\!=\!|100\rangle\!_m$ and
$|\psi\rangle\!_a^{(3)}\!\!\!=\!\!|0\rangle\!_{a\!_1}|0\rangle\!_{a\!_2}|0\rangle\!_{a\!_3}\!=\!|000\rangle\!_a$.
The single-excitation is set in the magnon 1. The magnon 1 is resonant with the cavity 1 by tuning the frequency of magnon 1, and the SC qubit is decoupled to the cavities. After time $T_1^{(3)}=\pi/2\lambda_{m_1}$, the local evolution $|1\rangle\!_{m\!_1}|0\rangle\!_{a\!_1}\rightarrow-\mathrm{i}|0\rangle\!_{m\!_1}|1\rangle\!_{a\!_1}$ is attained. The state is evolved to
\begin{eqnarray}    
|\psi\rangle_1^{(3)}=-\mathrm{i}|000\rangle\!_m\!|100\rangle\!_a|g\rangle\!_q.
\end{eqnarray}

The SC qubit is coupled to the three identical microwave cavities at the same time in far-detuning regime $\Delta_0\gg\lambda_q$. Therefore, the effective Hamiltonian of the subsystem composed of the SC qubit and the three identical cavities is of the form \cite{EH}
\begin{eqnarray}     
H_{\mathrm{eff}}^{(3)}&\!\!\!=\!\!\!&\widetilde{\lambda}_q\bigg[\sigma_z(a_1^\dagger a_1+a_2^\dagger a_2+a_3^\dagger a_3)+3|e\rangle_q\langle e|    \nonumber   \\
&&+\sigma_z(a_1a_2^\dagger+a_1a_3^\dagger+a_2a_3^\dagger+H.c.)\bigg].  \label{eff2}
\end{eqnarray}
The magnons are then all detuned with the cavities. The local evolution $e^{-\mathrm{i}H_{\mathrm{eff}}^{(3)}t}|100\rangle\!_a|g\rangle$ of the subsystem is given by
\begin{eqnarray}     
|\chi(t)\rangle\!_{\mathrm{sub}}^{(3)}&\!\!\!=\!\!\!&\bigg[C^{(3)}_{1,t}|100\rangle\!_a+C^{(3)}_{2,t}|010\rangle\!_a+C^{(3)}_{3,t}|001\rangle\!_a\bigg]   \nonumber   \\
&&\otimes|g\rangle\!_q,
\end{eqnarray}
where $C^{(3)}_{1,t}=\frac{e^{\mathrm{i}3\widetilde{\lambda}_qt}+2}{3}$ and $C^{(3)}_{2,t}=C^{(3)}_{3,t}=\frac{e^{\mathrm{i}3\widetilde{\lambda}_qt}-1}{3}$. It is easy to derive that
\begin{eqnarray}   
|C^{(3)}_{1,t}|^2+|C^{(3)}_{2,t}|^2+|C^{(3)}_{3,t}|^2=1.
\end{eqnarray}

\begin{figure}
\centering
\includegraphics[width=8.5cm,angle=0]{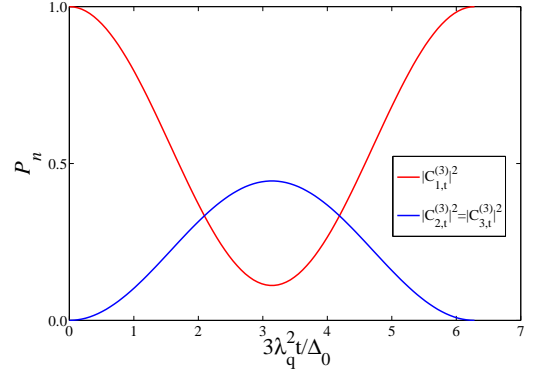}
\caption{(Color online) Evolution probabilities of the states: $P_1=|C^{(3)}_{1,t}|^2$ for $|100\rangle\!_a|000\rangle\!_m|g\rangle\!_q$ (red), $P_2=|C^{(3)}_{2,t}|^2$ for $|010\rangle\!_a|000\rangle\!_m|g\rangle\!_q$, $P_3=|C^{(3)}_{3,t}|^2$ for $|001\rangle\!_a|000\rangle\!_m|g\rangle\!_q$, and $P_2=P_3$ (blue). }\label{fig4}
\end{figure}

Fig.\ref{fig4} shows the probability related to the states $|100\rangle\!_a|000\rangle\!_m|g\rangle\!_q$, $|010\rangle\!_a|000\rangle\!_m|g\rangle\!_q$ and
$|001\rangle\!_a|000\rangle\!_m|g\rangle\!_q$. In particular, one has $|C^{(3)}_{1,t}|^2=|C^{(3)}_{2,t}|^2=|C^{(3)}_{3,t}|^2=\frac{1}{3}$, with
\begin{eqnarray}    
C^{(3)}_1=\frac{\sqrt{3}+\mathrm{i}}{2\sqrt{3}}, C^{(3)}_2=C^{(3)}_3=\frac{-\sqrt{3}+\mathrm{i}}{2\sqrt{3}}
\end{eqnarray}
at time $T_2^{(3)}=2\pi/9\widetilde{\lambda}_q$. Correspondingly, the state evolves to
\begin{eqnarray}     
|\psi\rangle_2^{(3)}&\!\!\!=\!\!\!&\bigg[\frac{\sqrt{3}+\mathrm{i}}{2\sqrt{3}}|100\rangle\!_a+\frac{-\sqrt{3}+\mathrm{i}}{2\sqrt{3}}|010\rangle\!_a
+\frac{-\sqrt{3}+\mathrm{i}}{2\sqrt{3}}|001\rangle\!_a\bigg]   \nonumber   \\
&&\otimes(-\mathrm{i})|000\rangle\!_m\otimes|g\rangle\!_q.
\end{eqnarray}

Finally, the magnons can be resonated with the respective cavities under the condition $\{\delta_1, \delta_2, \delta_3\}=0$. The local evolution and the time are
$|0\rangle\!_{m\!_k}|1\rangle\!_{a\!_k}\rightarrow-\mathrm{i}|1\rangle\!_{m\!_k}|0\rangle\!_{a\!_k}$ and $T_{3k}^{(3)}=\pi/2\lambda_{m_k}$ ($k=1, 2, 3$), respectively. Thus the final state is
\begin{eqnarray}      
|\psi\rangle_3^{(3)}&\!\!\!=\!\!\!&-\bigg[\frac{\sqrt{3}+\mathrm{i}}{2\sqrt{3}}|100\rangle\!_m\!+\!\frac{-\sqrt{3}+\mathrm{i}}{2\sqrt{3}}|010\rangle\!_m\!
+\!\frac{-\sqrt{3}+\mathrm{i}}{2\sqrt{3}}|001\rangle\!_m\!\bigg]   \nonumber   \\
&&\otimes|000\rangle\!_a\otimes|g\rangle\!_q.
\end{eqnarray}
In the whole process, the state of the SC qubit is kept unchanged.

\subsection{Numerical result}

\begin{figure}
\centering
\includegraphics[width=9.0cm,angle=0]{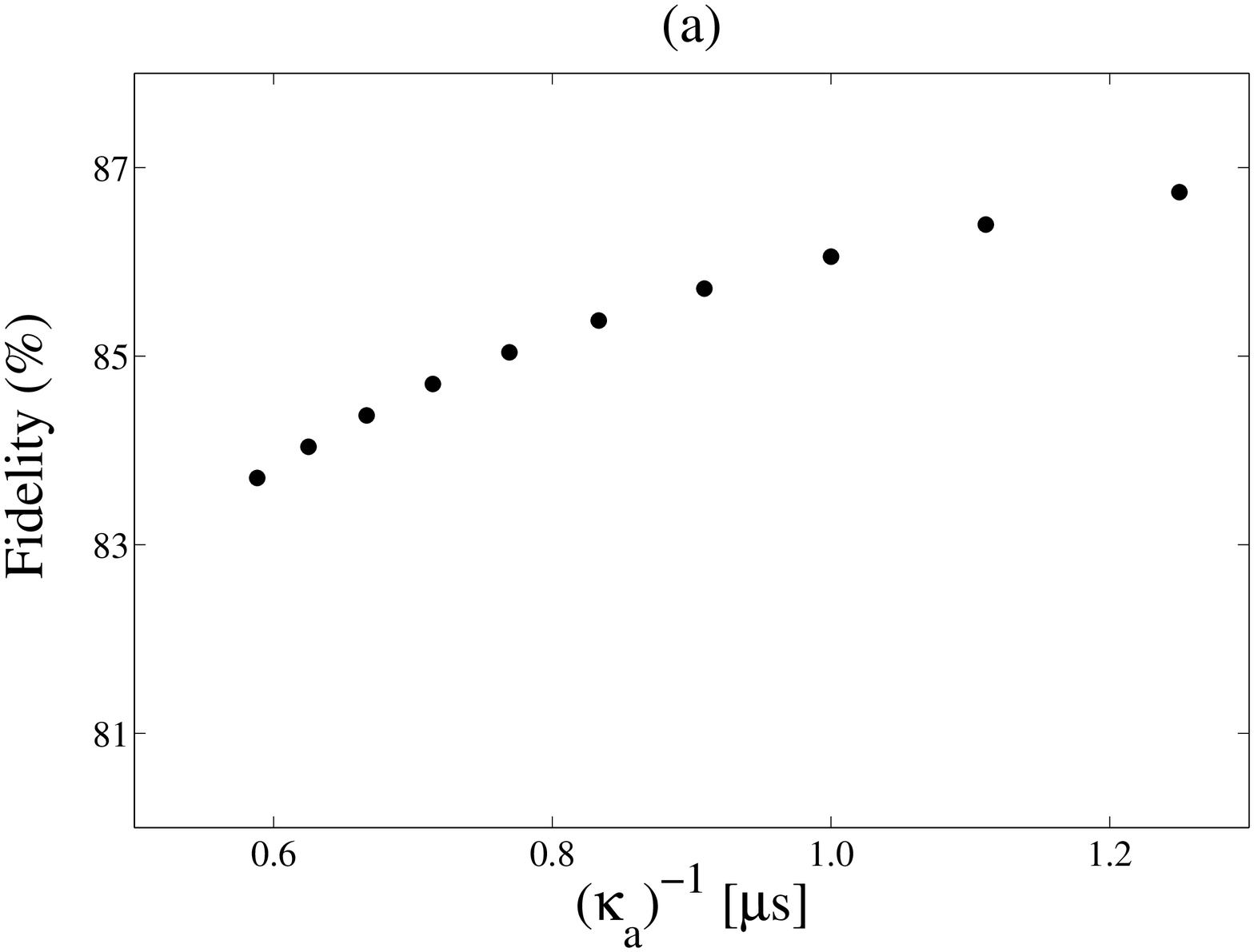}
\includegraphics[width=9.0cm,angle=0]{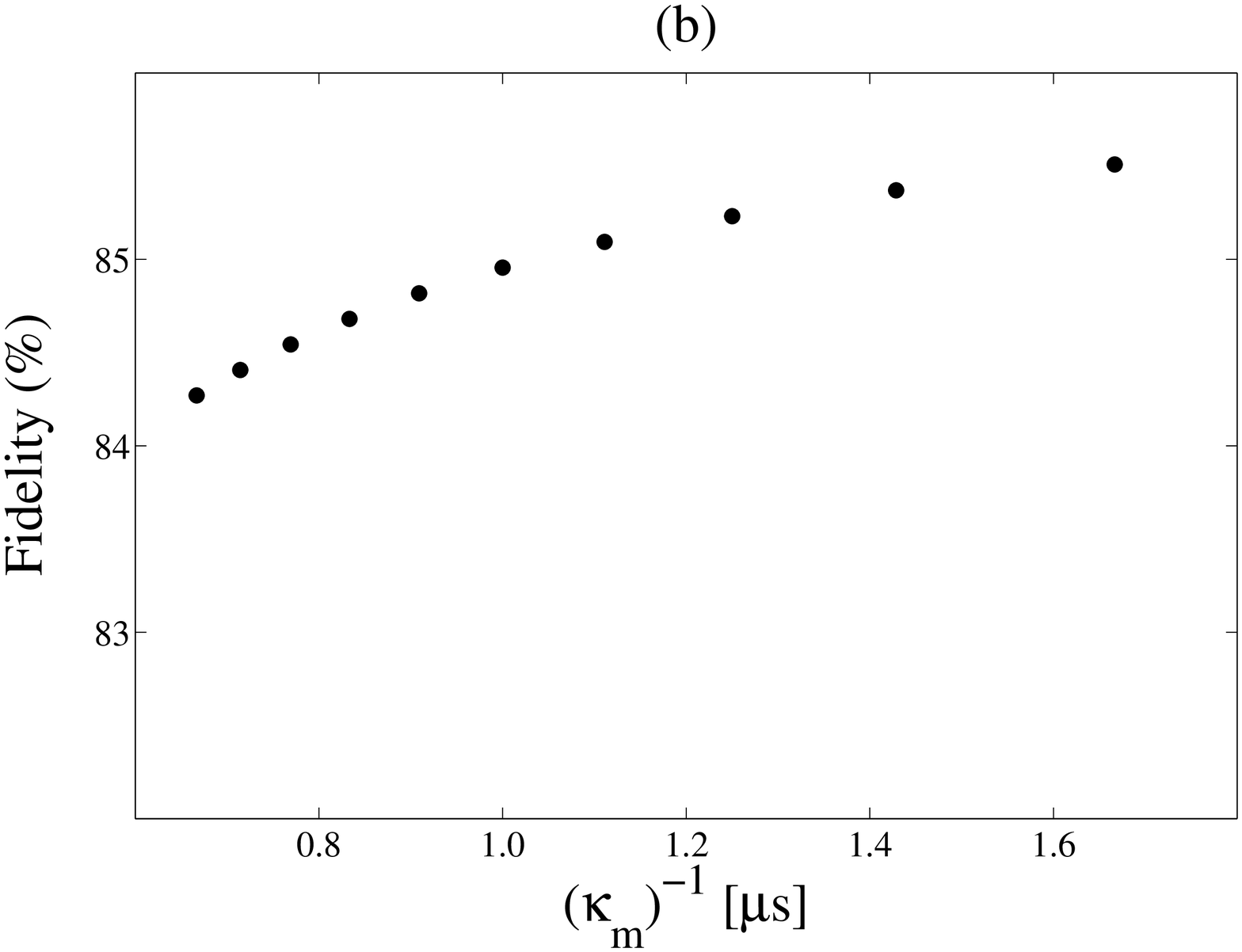}
\includegraphics[width=9.0cm,angle=0]{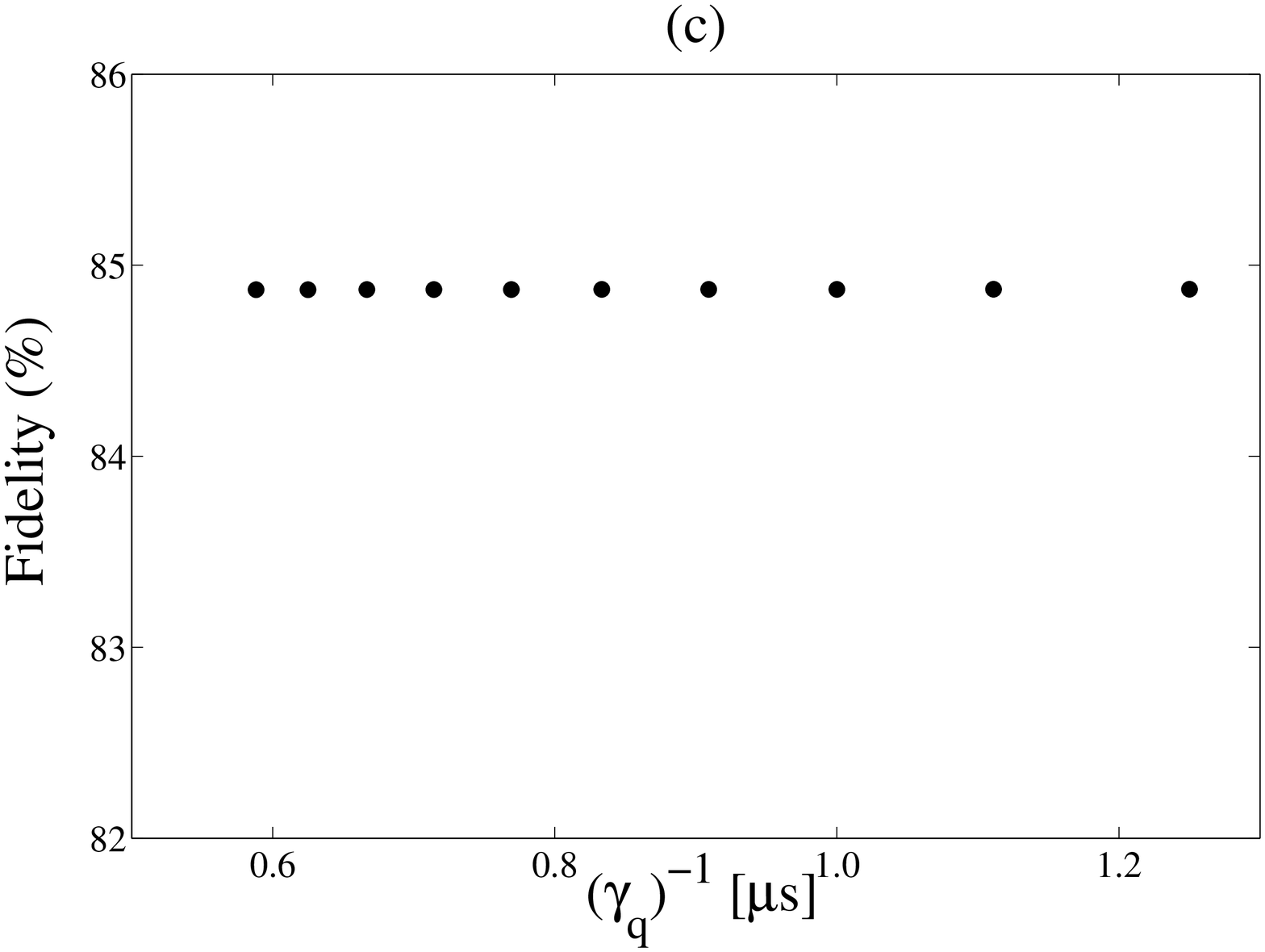}
\caption{ (a)-(c) The fidelity of the entanglement on three nonlocal magnons versus the dissipations of cavities, magnons and SC qubit.}\label{fig5}
\end{figure}

The entanglement fidelity of three nonlocal magnons is given here by taking into account the dissipations of hybrid system. Firstly, the master equation which governs the realistic evolution of the hybrid system composed of three magnons, three microwave cavities and a SC qubit can be expressed as
\begin{eqnarray}       
\dot{\rho}^{(3)}&\!\!=\!\!&-\mathrm{i}[H^{(I)}_3,\rho^{(3)}]+\kappa_{m_1}D[m_1]\rho^{(3)}+\kappa_{m_2}D[m_2]\rho^{(3)}    \nonumber   \\
&&+\kappa_{m_3}D[m_3]\rho^{(3)}+\kappa_{a_1}D[a_1]\rho^{(3)}+\kappa_{a_2}D[a_2]\rho^{(3)}    \nonumber  \\
&&+\kappa_{a_3}D[a_3]\rho^{(3)}+\gamma_qD[\sigma]\rho^{(3)},
\end{eqnarray}
where $\rho^{(3)}$ is the density operator of realistic evolution of the hybrid system, $\kappa_{m_3}$ is the dissipation rate of magnon 3 with $\kappa_{m_3}/2\pi=\kappa_m/2\pi=1.06$ MHz \cite{SU}, $\kappa_{a_3}$ denotes the dissipation rate for the microwave cavities 3 with $\kappa_{a_3}/2\pi=\kappa_a/2\pi=1.35$ MHz \cite{SU}, $D[X]\rho^{(3)}\!\!=\!\!(2X\rho^{(3)} X^\dag-X^\dag X\rho^{(3)}-\rho^{(3)} X^\dag X)/2$ for any $X=m_1, m_2, m_3, a_1, a_2, a_3, \sigma$.

The entanglement fidelity for three nonlocal magnons is defined by $F^{(3)}=_3^{(3)}\!\!\langle\psi|\rho^{(3)}|\psi\rangle_3^{(3)}$, which can reach 84.9$\%$. The fidelity with respect to the parameters is shown in Fig.\ref{fig5}.

\section{\emph{N} MAGNONS SITUATION}\label{4M}

\begin{figure}
\centering
\includegraphics[width=8.5cm,angle=0]{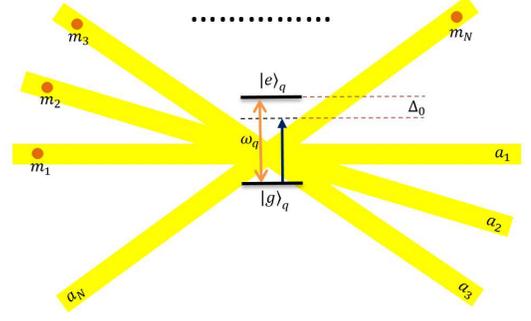}
\caption{(Color online) Schematic of the hybrid system composed of \emph{N} yttrium iron garnet spheres coupled to respective microwave cavities. A superconducting qubit is placed at the center of the \emph{N} identical microwave cavities. }\label{fig6}
\end{figure}

In Sec. \ref{2M} and Sec. \ref{3M}, the entanglement of two and three nonlocal magnons have been established. In this section we consider the case of \emph{N} magnons. In the hybrid system shown in Fig.\ref{fig6}, the SC qubit is coupled to \emph{N} cavity modes that have the same frequencies $\omega_a$. A magnon is coupled to the cavity mode in each cavity. Each magnon is placed at the antinode of microwave magnetic field of the respective cavity and biased static magnetic field.

In the interaction picture the Hamiltonian of whole system shown in Fig.\ref{fig6} can be expressed as
\begin{eqnarray}    
H^{(\mathrm{I})}_N\!\!&=\!\!&\sum_n \bigg[\lambda_{m_n}(a_nm^\dag_ne^{\mathrm{i}\delta_n t}+H.c.)    \nonumber  \\
&&\,\,\,\,\,\,\,\,\,\,\,\,\,+\lambda_q(a_n\sigma^+e^{\mathrm{i}\Delta_n t}+H.c.)\bigg],
\end{eqnarray}
where $a_n$ and $m^\dag_n$ ($n=1,2,3,\cdots,N$) are the annihilation operator of the \emph{n}th cavity mode and the creation operator of the \emph{n}th magnon, $\lambda_{m_n}$ is the coupling between the \emph{n}th magnon and the \emph{n}th cavity mode, $\lambda_q$ denotes the coupling strength between the SC qubit and the \emph{n}th cavity mode, $\delta_n=\omega_{m_n}-\omega_a$, $\omega_{m_n}$ is the frequency of the \emph{n}th magnon, $\Delta_n=\Delta_0=\omega_q-\omega_a$.

The initial state is prepared as
\begin{eqnarray}   
|\psi\rangle^{(N)}_0\!\!\!&=&\!\!\!|\psi\rangle\!_m^{(N)}\otimes|\psi\rangle\!_a^{(N)}\otimes|g\rangle\!_q,    \\
|\psi\rangle\!_m^{(N)}\!\!\!&=&\!\!\!|1\rangle\!_{m\!_1}|0\rangle\!_{m\!_2}|0\rangle\!_{m\!_3}\cdots|0\rangle\!_{m\!_N}\!=\!|100\cdots0\rangle\!_m,     \nonumber   \\
|\psi\rangle\!_a^{(N)}\!\!\!&=&\!\!\!|0\rangle\!_{a\!_1}|0\rangle\!_{a\!_2}|0\rangle\!_{a\!_3}\cdots|0\rangle\!_{a\!_N}\!=\!|000\cdots0\rangle\!_a.     \nonumber
\end{eqnarray}
At first, we tune the frequency of magnon 1 under the condition $\delta_1=0$. The magnon 1 is resonant with the cavity 1, which means that the single photon is transmitted to cavity 1, and the SC qubit is decoupled to all the cavities. The state evolves to
\begin{eqnarray}      
|\psi\rangle^{(N)}_1\!\!\!&=&\!\!\!-\mathrm{i}|000\cdots0\rangle\!_m|100\cdots0\rangle\!_a|g\rangle\!_q
\end{eqnarray}
after time $T_1^{(N)}=\pi/2\lambda_{m_1}$.

Next the magnons are tuned to detune with respective cavities. The SC qubit is coupled to the \emph{N} microwave cavities at the same time in far-detuning regime $\Delta_0\gg\lambda_q$. Under the condition $\Delta_n=\Delta_0$, the effective Hamiltonian of the subsystem composed of the SC qubit and \emph{N} microwave cavities is of the form \cite{EH}
\begin{eqnarray}       
H_{\mathrm{eff}}^{(N)}&\!\!=\!\!&\sum_n \widetilde{\lambda}_q\bigg[\sigma_za_n^\dagger a_n+|e\rangle_q\langle e|\bigg]    \nonumber   \\
&&+\sum_{l<n}\widetilde{\lambda}_q\bigg[\sigma_z(a_la_n^\dagger+H.c.)\bigg].
\end{eqnarray}
Consequently, the evolution of the hybrid system is given by
\begin{eqnarray}      
|\psi\rangle^{(N)}_2&\!\!\!=\!\!\!&\bigg[C^{(N)}_{1,t}|100\cdots0\rangle\!_a+C^{(N)}_{2,t}|010\cdots0\rangle\!_a   \nonumber   \\
&&+C^{(N)}_{3,t}|001\cdots0\rangle\!_a+\cdots+C^{(N)}_{N,t}|000\cdots1\rangle\!_a\bigg]     \nonumber    \\
&&\otimes(-\mathrm{i})|000\cdots0\rangle\!_m\otimes|g\rangle\!_q,
\end{eqnarray}
where
\begin{eqnarray}      
C^{(N)}_{1,t}&\!\!\!=\!\!\!&\frac{e^{\mathrm{i}N\widetilde{\lambda}_qt}+(N-1)}{N},      \nonumber  \\
C^{(N)}_{2,t}=C^{(N)}_{3,t}&\!\!\!=\!\!\!&\cdots=C^{(N)}_{N,t}=\frac{e^{\mathrm{i}N\widetilde{\lambda}_qt}-1}{N}. \label{coeff}
\end{eqnarray}
In addition, we have the following relation
\begin{eqnarray}      
\sum_n|C^{(N)}_{n,t}|^2\!\!&\!\!\!=\!\!\!&\!\!|C^{(N)}_{1,t}|^2+|C^{(N)}_{2,t}|^2+|C^{(N)}_{3,t}|^2+\cdots+|C^{(N)}_{N,t}|^2    \nonumber   \\
&=&\!\!1
\end{eqnarray}
by straightforward calculation.

At last, the SC qubit is decoupled to the cavities, and the magnons are resonant with the cavities, respectively. Thus, after the time $T_{3n}^{(N)}=\pi/2\lambda_{m_n}$, the final state is given by
\begin{eqnarray}     
|\psi\rangle^{(N)}_3&\!\!\!=\!\!\!&-\bigg[C^{(N)}_{1,t}|100\cdots0\rangle\!_m+C^{(N)}_{2,t}|010\cdots0\rangle\!_m   \nonumber   \\
&&+C^{(N)}_{3,t}|001\cdots0\rangle\!_m+\cdots+C^{(N)}_{N,t}|000\cdots1\rangle\!_m\bigg]     \nonumber    \\
&&\otimes|000\cdots0\rangle\!_a\otimes|g\rangle\!_q.
\end{eqnarray}
In the whole process, the state of SC qubit is unchanged all the time.

\begin{figure*}[!htbp]       
\centering
\begin{tabular}{ccc}
\includegraphics[width=5.5cm,angle=0]{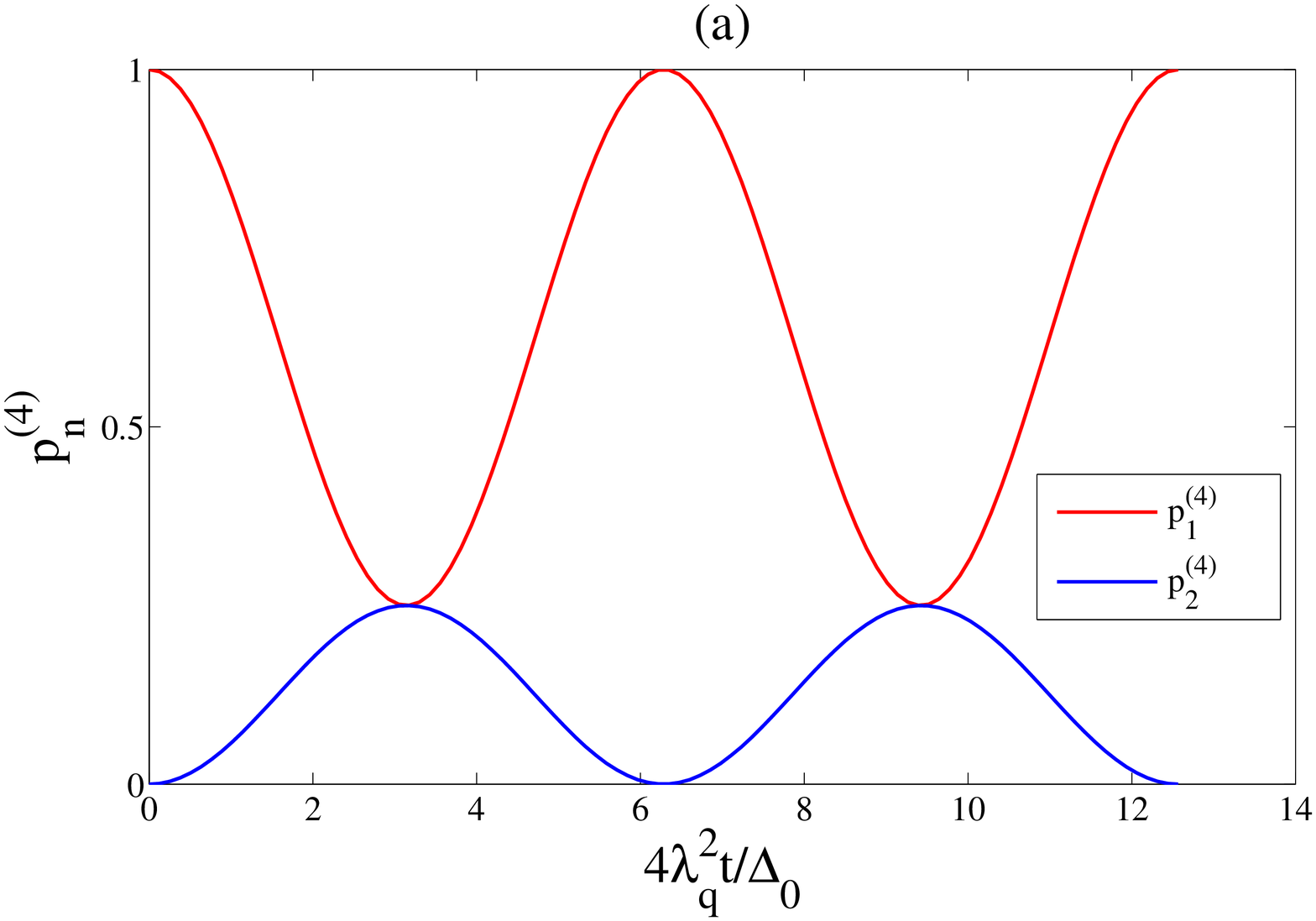} & \includegraphics[width=5.5cm,angle=0]{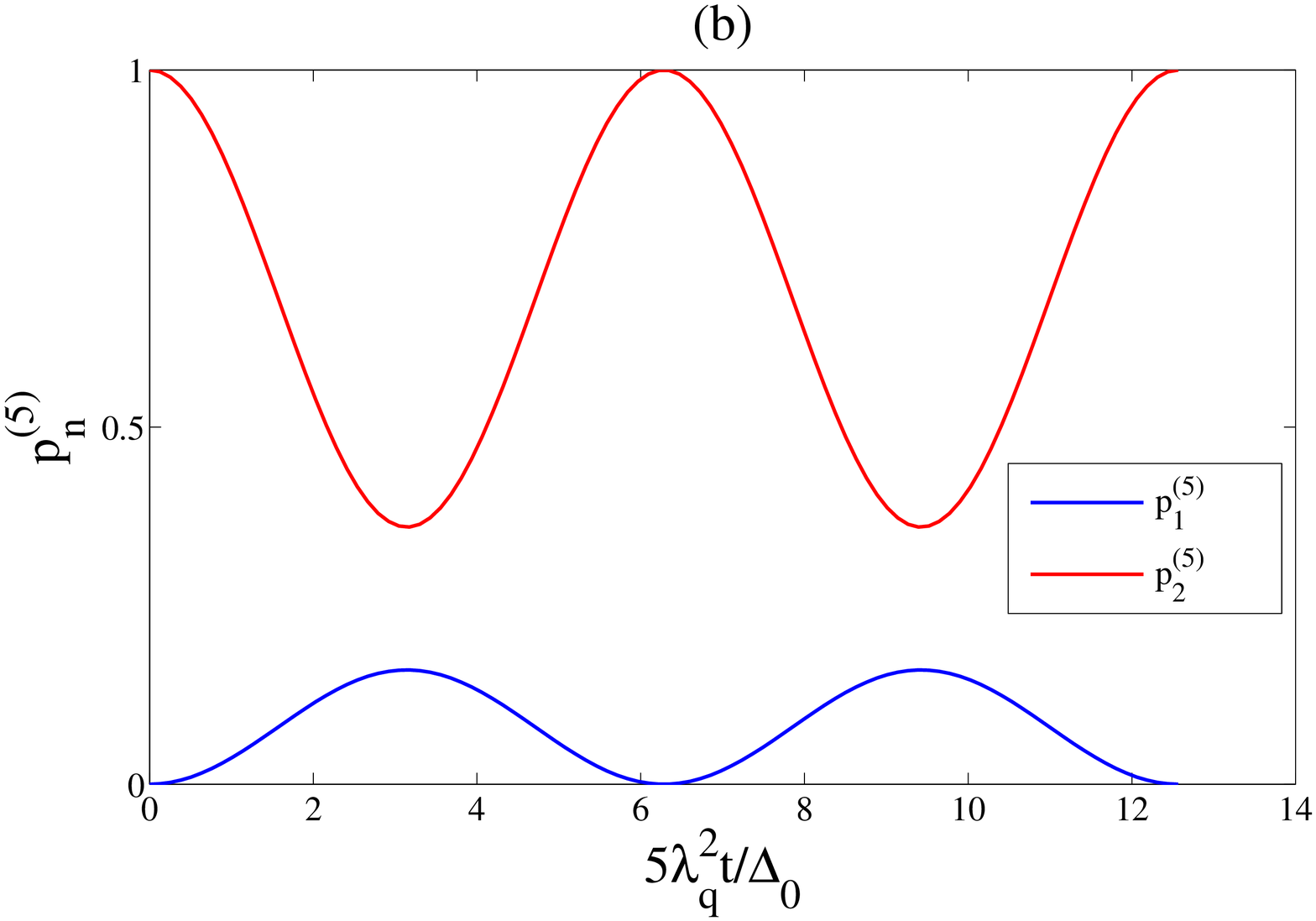} & \includegraphics[width=5.5cm,angle=0]{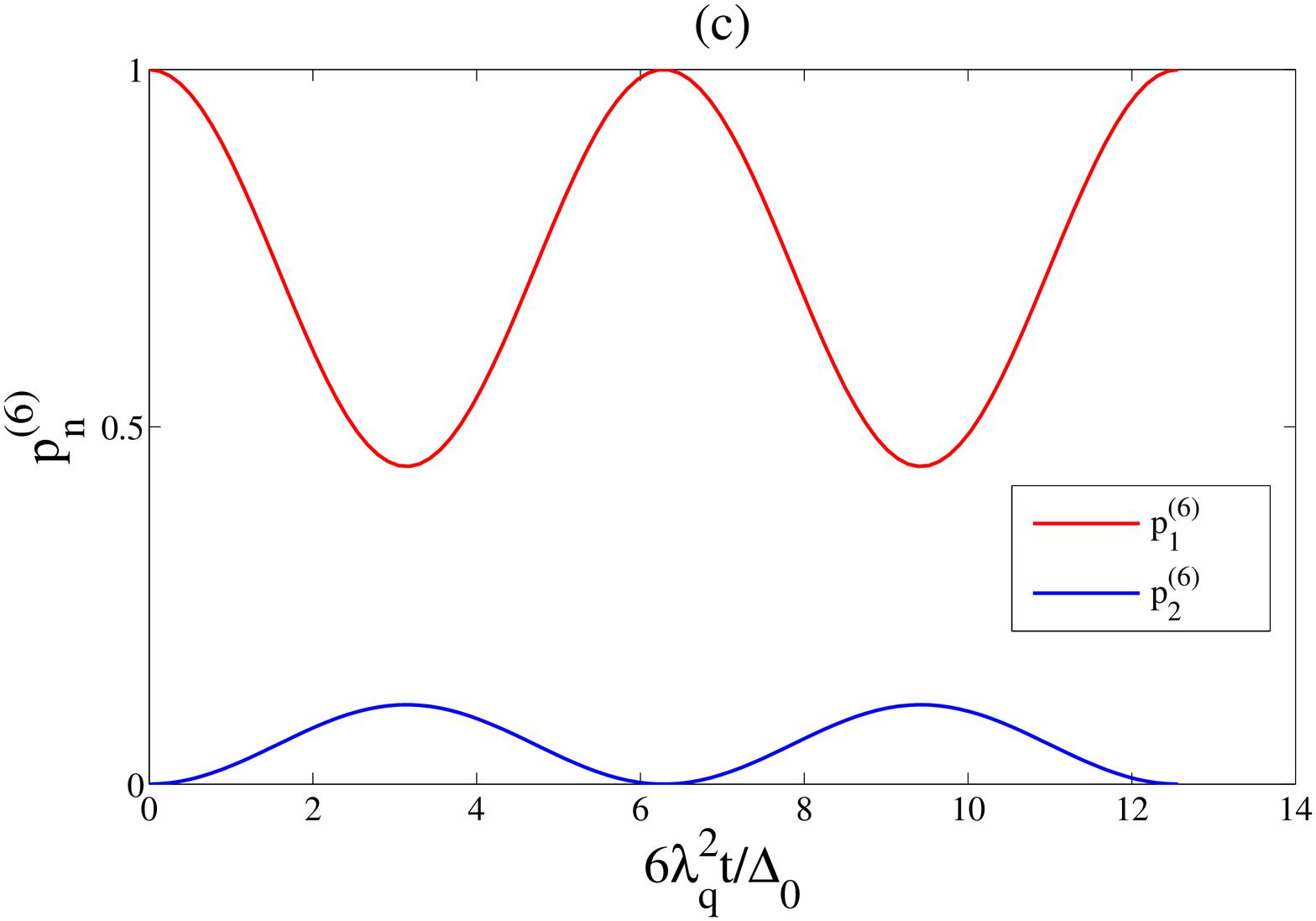}
\end{tabular}
\caption{(a)-(c) Evolution probabilities for $N=4$ (left), $N=5$ (middle), and $N=6$ (right). If $N\geqslant5$, $p_1^{(N)}\neq p_2^{(N)}$ implies that the isoprobability entanglement does not exist.}\label{fig7}
\end{figure*}

{\it [Remark]} Concerning the coefficients Eq. (\ref{coeff}), the probabilities with respect to the states $|100\cdots0\rangle\!_m|000\cdots0\rangle\!_a|g\rangle\!_q$, $|010\cdots0\rangle\!_m|000\cdots0\rangle\!_a|g\rangle\!_q$, $|001\cdots0\rangle\!_m|000\cdots0\rangle\!_a|g\rangle\!_q$, $\cdots$, $|000\cdots1\rangle\!_m|000\cdots0\rangle\!_a|g\rangle\!_q$ are $p^{(N)}_1=|C^{(N)}_{1,t}|^2$, $p^{(N)}_2=|C^{(N)}_{2,t}|^2$, $p^{(N)}_3=|C^{(N)}_{3,t}|^2$, $\cdots$, $p^{(N)}_N=|C^{(N)}_{N,t}|^2$, respectively, and $p^{(N)}_2=p^{(N)}_3=\cdots=p^{(N)}_N$. If the condition $p^{(N)}_1=p^{(N)}_2$ can be attained, the isoprobability entanglement can be obtained. For instance, for $N=4$, the entangled state of the four nonlocal magnons is given by
\begin{eqnarray}      
|\psi\rangle^{(4)}_3&\!\!\!=\!\!\!&-\frac{1}{2}\!\bigg[|1000\rangle\!_m\!-\!|0100\rangle\!_m\!-\!|0010\rangle\!_m\!-\!|0001\rangle\!_m\bigg]     \nonumber    \\
&&\otimes|0000\rangle\!_a\otimes|g\rangle\!_q.
\end{eqnarray}
However, if $N\geqslant5$, the isoprobability entanglement does not exist as a result of $p^{(N)}_1\neq p^{(N)}_2$, see illustration in Fig.\ref{fig7}(b)(c).

\section{SUMMARY AND DISCUSSION}\label{5M}

We have presented protocols of establishing entanglement on magnons in hybrid systems composed of YIGs, microwave cavities and a SC qubit. By exploiting the virtual photon, the microwave cavities can indirectly interact in far-detuning regime, and the frequencies of magnons can be tuned by the biased magnetic field, which leads to the resonant interaction between the magnons and the respective microwave cavities. We have constructed single-excitation entangled state on two and three nonlocal magnons, respectively, and the entanglement for \emph{N} magnons has been also derived in term of the protocol for three magnons.

By analyzing the coefficients in Eq. (\ref{coeff}), the isoprobability entanglement has been also constructed for cases $N=2$, $N=3$ and $N=4$. In particular, such isoprobability entanglement no longer exists for $N\geqslant5$.

In the protocol for the case of two magnons discussed in Sec. \ref{2M}, we have firstly constructed the superposition of magnon 1 and microwave cavity 1. Then the photon could be transmitted $|1\rangle\!_{a\!_1}|0\rangle\!_{a\!_2}\rightarrow-|0\rangle\!_{a\!_1}|1\rangle\!_{a\!_2}$ between two cavities. At last, the single-excitation Bell state is finally constructed in resonant way. As for $N\geqslant3$, however, such method is no longer applicable because of  $|100\cdots0\rangle\!_a\nrightarrow\alpha_2|010\cdots0\rangle\!_a
+\alpha_3|001\cdots0\rangle\!_a+\cdots+\alpha_N|000\cdots1\rangle\!_a$, namely, $p_1^{(N)}\neq0$.

\section*{Acknowledgements}

This work is supported by the National Natural Science Foundation of China (NSFC) under Grant Nos. 12075159 and 12171044, Beijing Natural Science Foundation (Grant No. Z190005), the Academician Innovation Platform of Hainan Province.

\end{document}